# Classification of Reverse-Engineered Class Diagram and Forward-Engineered Class Diagram using Machine Learning

Kaushil Mangaroliya

*Department of Computer Science and Engineering*

*Institute of Technology, Nirma University*

Ahmedabad, India

18bce091@nirmauni.ac.in

Het Patel

*Department of Computer Science and Engineering*

*Institute of Technology, Nirma University*

Ahmedabad, India

18bce076@nirmauni.ac.in

*Abstract* - UML Class diagram is very important to visualize the whole software we are working on and helps understand the whole system in the easiest way possible by showing the system classes, its attributes, methods, and relations with other objects. In the real world, there are two types of Class diagram engineers work with namely 1) Forward Engineered Class Diagram (FwCD) which are hand-made as part of the forward-looking development process, and 2). Reverse Engineered Class Diagram (RECD) which are those diagrams that are reverse engineered from the source code. In the software industry while working with new open software projects it is important to know which type of class diagram it is. Which UML diagram was used in a particular project is an important factor to be known? To solve this problem, we propose to build a classifier that can classify a UML diagram into FwCD or RECD. We propose to solve this problem by using a supervised Machine Learning technique. The approach in this involves analyzing the features that are useful in classifying class diagrams. Different Machine Learning models are used in this process and the Random Forest algorithm has proved to be the best out of all. Performance testing was done on 999 Class diagrams.

## I. INTRODUCTION

UML stands for Unified Modelling Language. It is used for general purpose, development, modeling language in the area of software engineering. It is a standard way of visualizing the design of the system. Software is almost very big most of the time. Making software can be a long and time-consuming process and having a clear plan about the complete process and the software is very important for every stakeholder involved. UML Class diagram is very important for all the developers because it helps them visualize what the ideas are to be implemented and understand what the client wants. Having a CML class diagram makes the implementation process easier and a lot fewer bugs and problems are expected at end of SDLC. UML class diagrams might not seem to be a big deal when working on a small project but when the software is huge, the UML class diagram is the most important so that stakeholders especially developers can understand what the ideas are and implementation is made easy.

SDLC comprises of many stages, some dependent and some independent of each other. In the beginning phases of the SDLC, class diagrams may be utilized to depict the architectural software design. As software starts to build, class diagrams can be utilized to depict information that is nearer to the development of the framework. While implementing or a bit after the implementation of source code, a class diagram might be recuperated with the help of some techniques. These types of class diagrams that are derived or reconstructed from the source code are called RWCD and they reflect small-small bits and pieces of the implementation structure of a software [1].

When it comes to applying machine learning to any problem, the main challenge to get data to solve that particular problem. Fortunately, Hebig et. al [2] present the Lindholmen dataset. It is a repository of all sorts of UML diagrams. It was built to fill in as an educational assortment of UML diagrams. This repository has more than 26,000 UML class charts and incorporates connections to the repositories on GitHub where the diagrams were found. As such it shapes a significant asset for observational investigations on ventures that utilization a few types of UML demonstrating. The research that has been conducted depends extensively on the availability of a large number of class diagrams. It has served its best in providing the data needed for the research. We contend that various acts of UML use, (example, utilization of forward and reverse engineered diagrams) can because of a cause of various impacts on different parts of programming frameworks. Be that as it may, these impacts are regularly ignored when it comes to a small bunch of programming projects. Classification of FwCD and RECD can prove to be of great benefit to the community. It can enable many researchers around the globe to make a case on the effect of these types of solutions on a bigger population in real-world technological industries.

This paper proposes a solution to the challenge discussed above. The solution in this paper might be helpful to the software industry. In this paper, a Supervised Machine Learning Classification Model has been proposed to accurately classify Forward-Engineered Diagram (FwCD) [3] and Reverse Engineered Class Diagram (RECD) [3]. We apply Supervised Machine Learning algorithms that have been chosen so that class diagrams can be given inputs along with their labels and the classifier can be trained accordingly. The dataset we have to utilize is a dataset of 999 class diagrams that are gathered from the Lindholmen dataset. To get a ground truth, this dataset is manually labeled by specialists that have involvement with working with UML diagrams and the software industry for years. The classification features are the most important parameters because the model is trained on these values and it behaves according to the features so the features need to be selected very carefully. The classification features are extricated after the guidance and view of specialists and their precise judgment. In this paper 11 machine learning classification algorithm was trained on the dataset. After training these 11 algorithms, an evaluation was done and the best algorithm was chosen.

Each study that is made by an individual or a group of an individual has some amount of contribution, small or big. This paper contributes by (I)identifying features for classification of class diagrams, (ii) robust supervised machine learning algorithm for class diagram classification, (iii) a dataset with ground truth for classification of class diagrams, (iv) analysis and comparison of the performance of various supervised machine learning algorithm on the problem.

## II. RELATED WORK

In recent times there has been no work that is extensively focused on the classification of class diagrams that is ground-breaking Due to this the search for related work was carried out further. There has been work done where class diagrams parameters and machine learning algorithms have been used for classification purposes.

One such work is by Maneerat and Muenchaisri [6]. In this paper, they discussed the method to predict bad-smell for the software design model. This paper took into consideration 27 standard software metrics which was proposed by Abreu in [7]. In one more paper in the same domain, Halim [8] wrote a paper that discussed the method to predict flaw inclined classes with help of the complexity metric of UML class diagrams. From various models in machine learning, two of the models proved to be good according to them, KNN and Naive Bayes. One more researcher Bagheri along with Gasevic [9] published a paper that discussed whether a set of structural metrics can prove to be a good predictor considering three main sub-characteristics of maintainability: analysability, changeability, and understandability.

One more research by Nugroho showcased a method to predict defects seeing the diagram metrics [5]. For this, he created measurements that were expected to catch the degree of detail of UML diagrams. Instances of the metrics that were proposed are the level of association-relations with labels, the rate of techniques that have marks. A conclusion was drawn that the higher the level of detailing in the UML models, the lesser will be the defects in the source code of the software. So, this led to researchers drawing one more conclusion that if the level of detail metrics is taken into consideration in any machine learning classifier then it can prove to be a highly accurate quality assurance model. A researcher Osman et. al [10] proposed a way to deal with RECD by utilizing machine learning models. An idea was proposed that the use of standard object-oriented design metrics (size measures and coupling measures) should be made to get better results. After this work, Thung et. al [11] worked on classification and made performance better by adding Network measure (for example Bay center, PageRank).

## III. METHODOLOGY

The process of solving the problem involves many steps such as data collection, feature extraction, and ground truth extraction, model learning, and getting the evaluation and results.

### A. Data Collection

Dataset is the first requirement when working with a machine learning model. In this paper, the use of UML class diagram images was used that was gathered from the Lindholmen dataset [2]. It was made possible by filtering 4443 software and gathered 2000 class diagram and were stored in different image file formats. The dataset needs to be clean and with minimal disturbance. Dataset cleaning was carried out and any duplicate images were removed. At last 999 class diagrams were derived which were perfect to be used with the model.

One problem to be solved before moving forward was that class diagrams were in image format. To be able for them too with machine learning algorithms, data in images is supposed to be extracted. This process of feature extraction was necessary to move forward. To solve this problem, XML Metadata Interchange was used. The idea is to convert Class diagrams from images to XMI format and this can be done with help of Image2UML tool. In this process, a couple of important points were noted which was that the tool could capture more information like the operator parameter.

### B. Feature Extraction & Ground Truth

Machine Learning models get trained on the labeled dataset. Supervised Machine Learning algorithms need to be labeled dataset for learning purposes. As there are no stated guidelines or instructions to recognize FwCD or RECD, this labeling was done physically by three chose UML specialists who have at any rate five years of involvement in utilizing UML class diagrams. The dataset included all this and was best suited for the paper. The labeling cycle comprised of two stages: (I) All the specialists assembled in a meeting to generate new ideas about the characteristics of FwCD and RECD. This was already done in the dataset (ii) Each person was haphazardly allocated a bunch of 323 class diagram. For each diagram, each person needs to group it into various classes: "FW Design", "RE Design" and "For Discussion". Diagrams in "For Discussion' were examined in the later stage. The cycle was carried out until all diagrams were classified. We were lucky to be able to get the dataset that had all this done at first.

In prior work of a fellow researcher Osman et. al [4], he mentioned a couple of characteristics that are common and repetitive of UML diagrams which were reverse-engineered using tools (by utilizing chose business

UML-CASE instruments). These characteristics ended up to be the final classification features. To this, highlights that the people from industry evoked when naming charts as FwCD and RECD were added.

*C. Model Training*

TABLE 1. FEATURES & INFOGAIN RESULTS

| Feature | Description | Information Gain |
|---|---|---|
| avgParaOper | Average parameter per operation | 0.3211 |
| numPara | The total number of parameters in the class diagram | 0.2830 |
| extOperPara | "true" - operation parameter exist "false"- not exist | 0.2410 |
| avgOperCls | Average operation per class | 0.2249 |
| maxOperCls | Select the highest number of operation (for a class) | 0.1651 |
| avgAssocCls | Average association per class | 0.1502 |
| numCls | The total number of classes in the class diagram | 0.1324 |
| numAssoc | The total number of associations in the class diagram | 0.1409 |
| numOper | The total number of operations in the class diagram | 0.1291 |
| numOrpCls | The total number of orphan classes in the class diagram | 0.8421 |
| avgOrpCls | Average orphan classes per class diagram | 0.0698 |
| avgAttrCls | Average attribute per class | 0.0624 |
| numAttr | The total number of attributes in the class | 0.0531 |
| maxAttrCls | Select the highest number of attributes (in a class) | 0.0378 |
| extOrpCls | "true" - orphan (unconnected) classes exist "false" - not exist | 0.0241 |
| numAssocType | Count the number of association type that exist in the class diagram | 0.0113 |

There are multiple algorithms and models under machine learning. Some algorithms work better on a particular problem while others don't, there is no particular algorithm that can be applied to all problems and for all purposes. The only way to know if a particular algorithm will be suitable or not is to experiment with a range of machine learning algorithms.

Different algorithms work differently, their output depends on the way they work hence a selection of proper algorithms is important. For instance, Decision Trees, Stumps, Tables, and Random Trees or Forests are some algorithms that work similarly, all divide the input into sub-parts, and try to make a prediction taking into consideration the occurrence of positive classes in sub-parts of input. K Nearest Neighbours and Radical Basis Function work in almost the same manner but unlike decision trees, their space overlaps. On the other hand, Regression is opposite to these algorithms because the model parameters are estimated considering a large number of data points and hence it is a more global model [10]. OneR is comparatively simple and is easy to understand classification algorithm. In this paper, the results of ZeroR give the probability of a class diagram being ReCD or FwCD. The feature pre-existing are more than relaiable. For evaluation of the models, K-Fold cross-validation gives a good insight. In this paper, stratified 10-fold cross-validation is used to evaluate the classification algorithm performance. Further 10-fold cross-validation is repeated 10 times for every classification model. WEKA tool makes this process easier.

*D. Evaluation of Results*

To evaluate the predictive power of predictors, the use of information gain wrt. the classes have been made. Univariate predictive power implies estimating how powerful a single indicator is in terms of prediction power. This particular technique has been applied in this paper. Use of WEKA's Information Gain Attribute Evaluator (InfogainAttrEval) has been made in carrying this test. The WEKA's InfogainAttrEval outputs a numeric value from 0 to 1. The higher estimation of InfoGain (near 1) means the predictor is more having more influence.

By and large, in this paper use of three evaluation measures to evaluate the performance has been made, for example (I) Percentage correctness (Accuracy), (ii) Precision and, (iii) Recall. Whenever required, the broadening of this assessment into more detail measures was done, for example, F-Measure and ROC-AUC Curve were used.

## IV. RESULTS

*A. RQ1: Analysis of Selected Features*

InfoGain is one of the parameters that let us know how well the model is. Table 1 shows InfoGain results for all features used. This outcome shows that every feature taken into consideration in this model has some predictive mower no matter what. The higher the information gain better the parameter is so, an average number of parameters (avgPaaraOper) is a good parameter. Similarly, all three most powerful parameters are related to operation parameters. It was noticed that the other most influential parameters are several operations per class (avgOperClas) and the maximum number of operations per class (maxOperCls). All the parameters with high impact are related to operations in the class diagram. This proves that one of the main classification factors if the class diagram is RECD or FwCD, is the operations in class diagrams.

## B. RQ2: Classification Model Performance

While working on the problem, ZeroR value was found to be 80.68 which conforms to dataset imbalance. It was noticed that majority of the diagrams were RECD in the dataset. When all classification models and their performance was tabulated, it was found that all classification models give a significantly better performance compared to ZeroR (Table 2).

TABLE 2. CLASSIFICATION PERFORMANCE

| Performance Measure | Accuracy | Precision | Recall | F Measure | AUC |
|---|---|---|---|---|---|
| OneR | 88.01 | 0.94 | 0.91 | 0.92 | 0.84 |
| Logistic Reg. | 88.90 | 0.94 | 0.93 | 0.93 | 0.94 |
| Naive Bayes | 88.10 | 0.97 | 0.88 | 0.92 | 0.91 |
| RBFNetwork | 88.32 | 0.95 | 0.91 | 0.93 | 0.92 |
| SVM | 87.28 | 0.88 | 0.97 | 0.93 | 0.71 |
| Decision Table | 88.33 | 0.94 | 0.91 | 0.92 | 0.84 |
| KNN - 5 | 89.16 | 0.93 | 0.93 | 0.93 | 0.88 |
| Decision Stump | 87.69 | 0.98 | 0.87 | 0.92 | 0.89 |
| Random Tree | 87.89 | 0.92 | 0.93 | 0.93 | 0.81 |
| Random Forest | 90.74 | 0.95 | 0.93 | 0.94 | 0.96 |

Another baseline to take into consideration is OneR. Working of OneR is a bit different, it takes into consideration only one feature to construct the model. A researcher Hilte proposed that it is definitely possible that an algorithm works well in a dataset hence the above measures have to be adopted. OneR is the simplest classification model. Hence complex models should be able to give significant improvement because more computational power is required in those models. The performance of SVM, KNN, Decision Stump, and Random Tree was lower than OneR. Hence these models were excluded because they were below the baseline.

After the baseline study, the focus was shifted to comparison based on precision and recall values. Decision Table, Logistic Regression, KNN, and Random Forest gave a better result and were suitable to the dataset. Random Forest gave the best performance when looked at F1 and AUC curve, hence Random Forest is the best suitable in solving the problem in this paper. Still, there is a scope of enhancement that can be done by reconfiguring the classification parameters, combining classification algorithms or ensemble learning, training on a larger dataset, and so on.

## V. CONCLUSION

There are a couple of conclusions that can be drawn out of this paper. The work done in this includes the presentation of construction and evaluation of multiple machine learning classifiers used for classifying FwCD and RECD. Different features of the class diagram were looked and the best features were chosen for the classifier and the classifier was trained on those features. Multiple algorithms performed better than the benchmark but Random Forest turned out to be the best algorithm.